\documentclass{article}

\usepackage[utf8]{inputenc}
\usepackage[english]{babel}
\usepackage[T1]{fontenc}

\usepackage{amsthm}
\usepackage{mathrsfs}
\usepackage{fancyhdr}
\usepackage[colorlinks,linkcolor=red,citecolor=blue,urlcolor=black,filecolor=blue,backref=page]{hyperref}
\usepackage{graphicx}
\usepackage{amsmath}
\usepackage{amssymb}
\usepackage{dsfont}
\usepackage{geometry}
\usepackage{setspace}

\normalfont\upshape

\usepackage{natbib}

\usepackage{fullpage}
\usepackage{authblk}

\newcommand{\prob}{\mathbb{P}}
\newcommand{\mean}{\mathbb{E}}

\newcommand{\Var}{\mathbb{V}}
\newcommand{\Sd}{\text{sd}}

\newcommand{\stdev}{\text{sd}}
\newcommand{\ud}{\mathop{}\!\mathrm{d}}
\newcommand{\cond}{\,|\,}

\newcommand{\e}{e} 

\DeclareMathOperator{\Normal}{\mathcal{N}}

\DeclareMathOperator{\Unif}{\mathcal{U}}

\newcommand{\Bs}{\mathbf{s}}

\newcommand{\Bx}{\mathbf{x}}

\newcommand{\Btheta}{\boldsymbol{\theta}}

\newcommand{\epsi}{\varepsilon}

\newcommand{\reals}{\mathbb{R}}
\newcommand{\realsp}{\mathbb{R}_{+}}

\newcommand{\X}{\mathscr{X}}
\newcommand*{\half}[1][2]{\frac{1}{#1}}

\newcommand{\eqdef}{\triangleq}

\newcommand{\indic}{\mathds{1}}

\newcommand{\Bxobs}{\Bx_{\textnormal{o}}}
\newcommand{\Bxobsi}{\Bx_{\textnormal{o},i}}
\newcommand{\Bsobs}{\Bs_{\textnormal{o}}}

\newcommand{\genpost}{\pi_{l}}
\newcommand{\genpostpr}{\pi_{l'}}
\newcommand{\genpostd}{\pi_{l_{\Delta}}}
\newcommand{\genpostgd}{\pi_{l_{g(\Delta)}}}
\newcommand{\pdf}{\pi}
\newcommand{\pdfabc}{\pdf_{\text{ABC}}}
\newcommand{\pdfabcerr}{\pdf_{\text{ABC,err}}}
\newcommand{\pdfabcl}{\pdf_{\text{ABC},l}}
\newcommand{\labc}{l_{\textnormal{ABC}}}
\newcommand{\truepdf}{\pdf^{*}}
\newcommand{\Edth}{\mean(\Delta_{\Btheta})}
\newcommand{\Vdth}{\Var(\Delta_{\Btheta})}

\newcommand{\Vd}{\sigma^2_{\!\Delta}}
\newcommand{\Edthstar}{\mean(\Delta_{\Btheta_{*}})}

\newcommand{\Sdthstar}{\stdev(\Delta_{\Btheta_{*}})}

\newcommand{\kernel}{weight function}

\renewcommand\cite{\citep}

\newtheorem{exmp}{Example}

\normalfont\upshape

\title{Surrogate-based ABC matches generalized Bayesian inference under specific discrepancy and kernel choices }

\date{}

\author[1]{Marko Järvenpää} 
\author[1,2,3]{Jukka Corander}
\author[4,*]{Henri Pesonen}
\affil[1]{Department of Biostatistics, University of Oslo, Oslo, Norway}
\affil[2]{Parasites and Microbes, Wellcome Sanger Institute, Hinxton, Cambridgeshire, UK}
\affil[3]{Helsinki Institute for Information Technology HIIT, Department of Mathematics and Statistics, University of Helsinki, Helsinki, Finland}
\affil[4]{Oslo Center for Biostatistics and Epidemiology, Oslo University Hospital, Oslo, Norway}
\affil[*]{Corresponding author: henri.e.pesonen@medisin.uio.no}

\begin{document}

\maketitle

\begin{abstract} 
Generalized Bayesian inference (GBI) is an alternative inference framework motivated by robustness to modeling errors, where a specific loss function is used to link the model parameters with observed data, instead of the log-likelihood used in standard Bayesian inference. Approximate Bayesian Computation (ABC) refers in turn to a family of methods approximating the posterior distribution via a discrepancy function between the observed and simulated data instead of using the likelihood. In this paper we discuss the connection between ABC and GBI, when the loss function is defined as an expected discrepancy between the observed and simulated data from the model under consideration. We show that the resulting generalized posterior corresponds to an ABC-posterior when the latter is obtained under a Gaussian process -based surrogate model. We illustrate the behavior of the approximations as a function of specific discrepancy and kernel choices to provide insights of the relationships between these different approximate inference paradigms.
\end{abstract}

\section{Introduction} \label{sec:intro}

Complex parametric statistical models are a major work horse in many fields of science and Bayesian inference is often used for them as it allows coherent estimation and uncertainty quantification of the parameters of such models based on combining the likelihood of data with prior beliefs. However, during the last 10 years a growing understanding about the undesirable behavior of posterior inference under model misspecification has emerged. In essence, the parameter estimates, their uncertainty quantification, or both, can become highly misleading when the amount of data increases under a misspecified model, which calls for alternative and more robust methods of inference. For a general discussion about failure of Bayesian inference under misspecification and various remedies to improve it, we refer the reader to ~\citet{Yang2007, Walker2013,Bissiri2016,Grunwald2017,Jewson2018,Yang2018}. 

A particular choice is to update the prior without directly using the likelihood, in order to avoid badly calibrated and thus misleading posterior distribution. In generalized Bayesian inference (GBI) one uses a loss function to update the prior, such that the loss function is designed to specifically measure fidelity against aspects of data that matter most to the considered application of the model \citep{Bissiri2016}. Generalized Bayesian inference of this form has been extensively investigated both theoretically and empirically by \citet{Jewson2018,Knoblauch2022,Matsubara2022,Frazier2022lossbased,Wu2023} among others.

An important but less acknowledged motivation for GBI is computational, since the likelihood function of a complex statistical model may be unavailable or very costly to evaluate. 
This computational difficulty can in principle be bypassed by various likelihood-free inference methods, such as Approximate Bayesian Computation (ABC), see \citet{Sisson2019}, where model simulations are used to replace the likelihood in an implicit or explicit manner. However, estimating a loss function as done in GBI can sometimes be more straightforward and computationally efficient as discussed in \citet{Pacchiardi2021,Gao2023} and \citet{Jarvenpaa2023}.  Robustness to model misspecification can then be obtained as a by-product of the GBI process.

Our main contribution in this article is to scrutinize the relationship between GBI and ABC, and in particular to show that under certain assumptions the generalized posterior is closely related to the posterior obtained under ABC with some special kernel and distance functions. We then discuss the practical relevance of this connection especially for Gaussian process-based surrogate methods for sample-efficient likelihood-free inference developed by \citet{Gutmann2016,Jarvenpaa2018_acq,Jarvenpaa2020_babc,Thomas2023,Jarvenpaa2023} and recently demonstrated in applied setting by \citet{Pesonen2023}.

\section{Generalized Bayesian inference} \label{sec:gbi}

In Bayesian inference the prior density $\pdf(\Btheta)$ for the model parameters $\Btheta\in\Theta\subset\reals^p$ is updated to the posterior distribution $\pdf(\Btheta\cond\Bxobs) = \pdf(\Btheta)\pdf(\Bxobs\cond\Btheta)/\pdf(\Bxobs)\propto \pdf(\Btheta)\pdf(\Bxobs\cond\Btheta)$ using the data $\Bxobs\in\X\subset\reals^d$. The updating crucially depends on the likelihood function $\pdf(\Bxobs\cond\Btheta)$ defined by the assumed model. 

However, when the statistical model is misspecified so that the likelihood $\pdf(\Bxobs\cond\Btheta)$ is at best a crude approximation to the underlying data generating process $\truepdf(\Bxobs)$ no matter how $\Btheta$ is chosen, the standard posterior may not lead to sensible parameter estimates or meaningful uncertainty quantification. To remedy this, \citet{Bissiri2016} discusses a generic, principled approach where the prior is instead updated via a loss function $l:\Theta\times\X\rightarrow\reals$ and which results in a generalized posterior
\begin{align}
    \genpost(\Btheta\cond\Bxobs) \propto \pdf(\Btheta)\e^{-wl(\Btheta,\Bxobs)}, \label{eq:genpost}
\end{align}
where $w>0$ is a scaling constant (sometimes included directly to the loss $l$). The loss function can be tailored for the inference problem at hand and selecting $w=1$ and $l(\Btheta,\Bxobs)=-\log\pdf(\Bxobs\cond\Btheta)$ reproduces the standard posterior $\pdf(\Btheta\cond\Bxobs)$. 

Here we focus on the ABC inference setting and consider the generalized posterior $\genpostd(\Btheta\cond\Bxobs)$ based on the loss function
\begin{align}
    l_{\Delta}(\Btheta,\Bxobs) 
    \eqdef \mean_{\Bx\cond\Btheta}\Delta(\Bxobs,\Bx) 
    = \int \Delta(\Bxobs,\Bx) \pdf(\Bx\cond\Btheta)\ud\Bx, \label{eq:meandiscrloss}
\end{align}
where $\Delta(\Bxobs,\Bx)$ quantifies some notion of distance (or similarity) between the observed data $\Bxobs$ and synthetic data $\Bx$ simulated from the model. 
Target densities of this form are used in \citet{Thomas2023,Gao2023} and also appear in \citet{Oliveira2021} and \citet{Jarvenpaa2023} but we find that no clear justification or interpretation were given there. Also, \citet{Pacchiardi2021} proposed a framework for generalized Bayesian likelihood-free inference where the loss is formed as a scoring rule and is of the form \eqref{eq:meandiscrloss} in some cases. 
%


\section{Model-based ABC as generalized Bayesian inference} \label{sec:main_analysis}


We start by presenting the common ABC target posterior $\pdfabc(\Btheta \cond \Bxobs)$ with our later analysis in mind. 
It is defined as $\pdfabc(\Btheta \cond \Bxobs) \propto \pdf(\Btheta) \pdfabc(\Bxobs \cond \Btheta)$ where
\begin{align}
    \pdfabc(\Bxobs \cond \Btheta) 
    \eqdef \mean_{\Bx\cond\Btheta}K_h(\Delta(\Bxobs,\Bx)) 
    = \int K_h(\Delta(\Bxobs,\Bx))\pdf(\Bx\cond\Btheta)\ud\Bx \label{eq:abc_lik}
\end{align}
is the ABC likelihood. Above $\Delta: \X^2\rightarrow\realsp$ denotes the distance and $K_h: \reals\rightarrow\realsp$ is a kernel function with bandwidth $h>0$. 
Many sampling algorithms have been proposed that target $\pdfabc(\Btheta \cond \Bxobs)$ and only require simulations from $\pdf(\Bx\cond\Btheta)$. Here we are however mainly interested understanding the approximate target posterior. 

In the context of ABC, $\Delta$ is often a discrepancy function and defined to compare summary statistics $S:\X\rightarrow\reals^s$ due to reasons of computational efficiency. That is, $\Bxobs$ and $\Bx$ in  \eqref{eq:abc_lik} are then replaced by $\Bsobs=S(\Bxobs)$ and $\Bs=S(\Bx)$, respectively. The discrepancy can be some norm so that $\Delta(\Bxobs,\Bx) = ||\Bxobs-\Bx||$ though other choices are feasible and $\Delta(\Bxobs,\Bx)$ doesn't necessarily need to satisfy the mathematical properties of the distance but only provide some notion of similarity of the data sets. 
In this paper we do not take a stand how the discrepancy should be formed and we always use $\Bxobs$ and $\Bx$ to denote the data sets, summarized or not.

The kernel $K_h(r)=K(r/h)/h$ in  \eqref{eq:abc_lik} further controls the approximation. 
The usual choice is the uniform kernel $K_h(r)\propto\indic_{|r|\leq h}$ but any function that is symmetric so that $K(r)=K(-r)$, non-negative and satisfies the conditions $\int K(r)\ud r=1$, $\int rK(r)\ud r=0$ and $\int r^2 K(r)\ud r<\infty$ as mentioned in \citet[Section~1.6]{Sisson2019} could be used in principle. This definition is actually only used in $r \geq 0$ as the discrepancy is typically formed to be non-negative. A relevant property is also that $K_h(r)$ should converge to a point mass at $r=0$ denoted by $\delta_{0}(r)$ as $h\rightarrow 0$.  
Later we encounter kernels which do not satisfy all of these properties but still result interpretable posterior approximations. We will refer to such ``kernels'' as \kernel{}s.
%
In fact, the kernel can often be interpreted as an observation (or model) error, as pointed out by \citet{Wilkinson2013}. In principle, the kernel can also be replaced with a general probabilistic model $\pdf(\Bxobs\cond\Bx)$ in which case the target density becomes  
\begin{align}
    \pdfabcerr(\Bxobs \cond \Btheta) 
    = \int \pdf(\Bxobs\cond\Bx) \pdf(\Bx\cond\Btheta) \ud\Bx \label{eq:abc_lik_errmodel}
\end{align}
which is the exact posterior under the assumption of error model $\pdf(\Bxobs\cond\Bx)$. 


We rewrite  \eqref{eq:abc_lik} as $\pdfabc(\Bxobs \cond \Btheta) = \mean_{\Delta_{\Btheta}}K_h(\Delta_{\Btheta})$ where the distance $\Delta_{\Btheta}$ is now treated as a stochastic process whose randomness is induced by the intractable model $\pdf(\Bx\cond\Btheta)$. The stochastic process $\Delta_{\Btheta}$ is indexed by $\Btheta\in\Theta$ and has independent marginals. The dependence of $\Delta_{\Btheta}$ on the fixed observed data $\Bxobs$ is suppressed for brevity. 
The ABC posterior is formally obtained from $\genpost(\Btheta\cond\Bxobs)$ when $w=1$ and  
\begin{align}
l(\Btheta,\Bxobs) 
= \labc(\Btheta,\Bxobs) 
\eqdef -\log \pdfabc(\Bxobs\cond\Btheta)
= -\log\mean_{\Delta_{\Btheta}} K_h(\Delta_{\Btheta}). 
\label{eq:abcloss}
\end{align}
As already observed in \citet{Jarvenpaa2023}, if $K_h$ is Exponential kernel so that $K_h(r)\propto\e^{-|r|/h}$, then by Jensen's inequality we have $\labc(\Btheta,\Bxobs) \leq l_{\Delta}(\Btheta,\Bxobs)/h + c$, where $c$ is a constant with respect to $\Btheta$. A generalized posterior $\genpostd(\Btheta\cond\Bxobs)$ is hence obtained in this case when the ``ABC loss'' $\labc(\Btheta,\Bxobs)$, that gives  the ABC posterior $\pdfabc(\Btheta \cond \Bxobs)$, is upper bounded with the loss $wl_{\Delta}(\Btheta,\Bxobs)$ where $w=1/h$. Note that here and in the following we implicitly neglect all additive terms in the loss function that are constants with respect to $\Btheta$ as they do not affect the resulting generalized posterior due to the normalization. 

In the following we show that similar relationships as above hold -- but with (approximate) equality in the place of the inequality -- in some special situations of interest. 
Specifically, we look for a \kernel{} $K_h(r)$ so that the integral equation
\begin{align}
    -w\mean_{\Delta_{\Btheta}}(\Delta_{\Btheta}) = \log\int K_h(\Delta_{\Btheta}) \pdf(\Delta_{\Btheta}) \ud\Delta_{\Btheta} \label{eq:inteq} + c
\end{align}
%
is satisfied up to an additive constant $c$ with respect to $\Btheta$. In this case the generalized posterior $\genpostd(\Btheta\cond\Bxobs)$ and the ABC target density $\pdfabc(\Btheta \cond \Bxobs)$ formally coincide. 
In principle, $K_h(r)$ could be allowed to depend also on $\Btheta$ in  \eqref{eq:inteq}. 

We first suppose $\pdf(\Delta_{\Btheta}) = \delta_{\Delta^*_{\Btheta}}(\Delta_{\Btheta})$, where $\Delta^*_{\Btheta}$ is a fixed function $\Theta\rightarrow\realsp$. This condition can be assumed to hold asymptotically in the limit of infinite data in i.i.d.~setting with suitable discrepancy, though we do not analyse this further here. The condition also holds when the model is actually deterministic in the sense that it always outputs $\Bx^*_{\Btheta}\in\X$ when simulated with $\Btheta\in\Theta$ so that $\pdf(\Bx\cond\Btheta) = \delta_{\Bx^*_{\Btheta}}(\Bx)$. Then \eqref{eq:inteq} takes the form $-w\Delta^*_{\Btheta} = \log K_h(\Delta^*_{\Btheta}) + c$ which is clearly satisfied by the Exponential kernel $K_h(r)=e^{-c} \e^{-|r|/h}$ and $w=1/h$. 
That is, the generalized posterior $\genpostd(\Btheta\cond\Bxobs)$ with $w=1/h$ is equal to the ABC posterior based on the Exponential kernel with bandwidth $h$ in this special case. Furthermore, the Exponential kernel can here be interpreted as a Exponential error model for the otherwise deterministic model $\pdf(\Bx\cond\Btheta) = \delta_{\Bx^*_{\Btheta}}(\Bx)$. 
If $\pdf(\Btheta)\propto \indic_{\Btheta\in\Theta}$, then any parameter minimizing $\Delta^*_{\Btheta}$ maximizes the corresponding generalized posterior and the value of $w$ solely controls the posterior dispersion. If $h\rightarrow 0$, then the Exponential kernel (or error model) becomes increasingly accurate and the generalized posterior intuitively concentrates to $\arg\min_{\Btheta\in\Theta}\Delta^*_{\Btheta}$.

\subsection{ABC as GBI under modeling assumptions} \label{subsec:gaus_discr_analysis}

In this section we assume that the induced distribution of $\Delta_{\Btheta}$ is Gaussian at each $\Btheta\in\Theta$. 
That is, we base our analysis here on the assumption  
%
\begin{equation}
    \pdf(\Delta_{\Btheta}) = \Normal(\Edth,\Vdth), \label{eq:gaus}
\end{equation}
where the dependency on $\Bxobs$ is again suppressed. 
While this assumption does not hold exactly in practice as the distance should be formed to be non-negative, the central limit theorem can be used to justify  \eqref{eq:gaus} as a sensible approximation in the context of ABC, see \citet[Section~A.2]{Jarvenpaa2020_babc} for details. 
Furthermore, the tail probability $\prob(\Delta_{\Btheta} < 0)$ based on any meaningfully fitted Gaussian density for $\Delta_{\Btheta}$ is often negligible at any $\Btheta$ especially when the simulator-based model is misspecified. 
We also introduce a workaround to the assumption of  \eqref{eq:gaus} later in Section \ref{sec:add_analysis}. 

If the kernel is uniform so that $K_h(r)\propto\indic_{|r|\leq h}$ and if $\Delta_{\Btheta} \geq 0$, we have $\mean_{\Delta_{\Btheta}} K_h(\Delta_{\Btheta}) \propto \mean_{\Delta_{\Btheta}}\indic_{\Delta_{\Btheta}\leq h} = \prob(\Delta_{\Btheta}\leq h)$. On the other hand, if we assume  \eqref{eq:gaus} and redefine the uniform kernel as a \kernel{} $K_h(r)\propto\indic_{r\leq h}$ (so that it is no longer symmetric or integrable) to acknowledge that the distance can now technically be also negative, we obtain
\begin{equation}
    \labc(\Btheta,\Bxobs) = -\log \prob(\Delta_{\Btheta}\leq h) = -\log\Phi\left((h-\Edth)/\sqrt{\Vdth}\right), \label{eq:log_gauss_abc_loss}
\end{equation}
%
where $\Phi(\cdot)$ denotes the CDF of standard Gaussian density. Recall that we implicitly neglect additive constants in the loss functions. 

A more interesting result is obtained by considering the Exponential \kernel{} $K_h(r)\propto\e^{-r/h}$ which is also no longer symmetric or integrable. 
By using the well-known formula for the moment generating function of the Gaussian density at $-1/h$, we then obtain
\begin{align}
    \labc(\Btheta,\Bxobs) 
    = -\log\mean_{\Delta_{\Btheta}} \e^{-\Delta_{\Btheta}/h}
    = \frac{\Edth}{h} - \frac{\Vdth}{2h^2}
    = \frac{1}{h}\left[\Edth - \frac{\Vdth}{2h}\right]. 
    \label{eq:exploss1}
\end{align}
%
If we additionally assume that $\Vdth$ is constant with respect to $\Btheta$, the second term of  \eqref{eq:exploss1} can be neglected and we can write
%
\begin{equation}\label{eq:exploss1_homogsigma}
    \labc(\Btheta,\Bxobs) = \frac{1}{h}\Edth = \frac{1}{h}l_{\Delta}(\Btheta,\Bxobs).
\end{equation}
%
Hence, if the discrepancy is Gaussian distributed with constant variance, the ABC posterior based on the Exponential \kernel{} is equal to the generalized posterior $\genpostd(\Btheta\cond\Bxobs)$ with $w=1/h$. 
If the constant variance assumption is violated, then this equivalence holds with loss in  \eqref{eq:exploss1}. 

Finally, consider the Gaussian \kernel{} $K_h(r)\propto\e^{-(r-m_h)^2/(2\sigma_h^2)}$ where $h=\sigma_h$ and also $m_h\in\reals$ can be used to control the \kernel{}. The Gaussian \kernel{} is integrable so it can be interpreted as an error model. It is also symmetric if $m_h=0$. Using a Gaussian identity we obtain
\begin{align}
    \labc(\Btheta,\Bxobs) 
    &= -\log\int_{-\infty}^{\infty} \e^{-(\Delta_{\Btheta}-m_h)^2/(2\sigma_h^2)} \Normal(\Delta_{\Btheta} \cond \Edth,\Vdth) \ud\Delta_{\Btheta} \\
    &= -\log \Normal(\Edth \cond m_h,\Vdth+\sigma_h^2) \\
    &= \half\log(\Vdth+\sigma_h^2) + \half\frac{(\Edth-m_h)^2}{\Vdth+\sigma_h^2}, \label{eq:gausloss1}
\end{align}
where the equalities again hold up to additive constants. 

Assuming further that $\Vdth$ is constant so that $\Vd \eqdef \Vdth$ for all $\Btheta$, setting $m_h=0$ or, equivalently, replacing $\Delta_{\Btheta}$ with $\Delta_{\Btheta} + m_h$, and using the fact $\Vdth=\mean(\Delta_{\Btheta}^2) - \Edth^2$, we obtain 
\begin{align}
    \labc(\Btheta,\Bxobs) = \frac{1}{2(\Vd+\sigma_h^2)}\mean(\Delta_{\Btheta}^2). \label{eq:gausloss2}
\end{align}
We hence observe that if the discrepancy $\Delta_{\Btheta}$ is Gaussian distributed with constant variance $\Vd$ and if the Gaussian \kernel{} is used, then the corresponding ABC posterior equals the generalized posterior with loss $l_{\Delta^2}(\Btheta,\Bxobs)=\mean_{\Bx\cond\Btheta}(\Delta(\Bxobs,\Bx)^2)=\mean(\Delta_{\Btheta}^2)$ and $w=1/(2(\Vd+\sigma_h^2))$. 
If the constant variance assumption is violated, then the above equivalence holds with the more complicated loss  \eqref{eq:gausloss1}.

\begin{exmp}\label{example:1}
To illustrate the differences between ABC and GBI posteriors given kernel modeling assumptions, we consider a simple Gaussian model with linearly dependent heterogeneous noise. The considered model is 
$\Bx \mid \Btheta \thicksim \Normal(\Btheta, 0.2 \cdot \Btheta + 0.01)$ with $\Btheta \thicksim \Unif[0,10]$. Additionally we use a discrepancy function $\Delta(\Bxobs, \Bx) = |\Bxobs - \Bx|$, and assume observed data $\Bxobs = 3$. We set $h = 0.2$ and illustrate the approximate posterior distributions calculated via the assumptions made in the Section \ref{subsec:gaus_discr_analysis}. We calculate the ABC posterior via rejection ABC as a baseline measure and calculate approximate posterior ABC/GBI posteriors based on  \eqref{eq:log_gauss_abc_loss},  \eqref{eq:exploss1},  \eqref{eq:exploss1_homogsigma},  \eqref{eq:gausloss1} and \eqref{eq:gausloss2}.

\begin{figure}[hbtp] 
\centering
\includegraphics[width=0.8\textwidth]{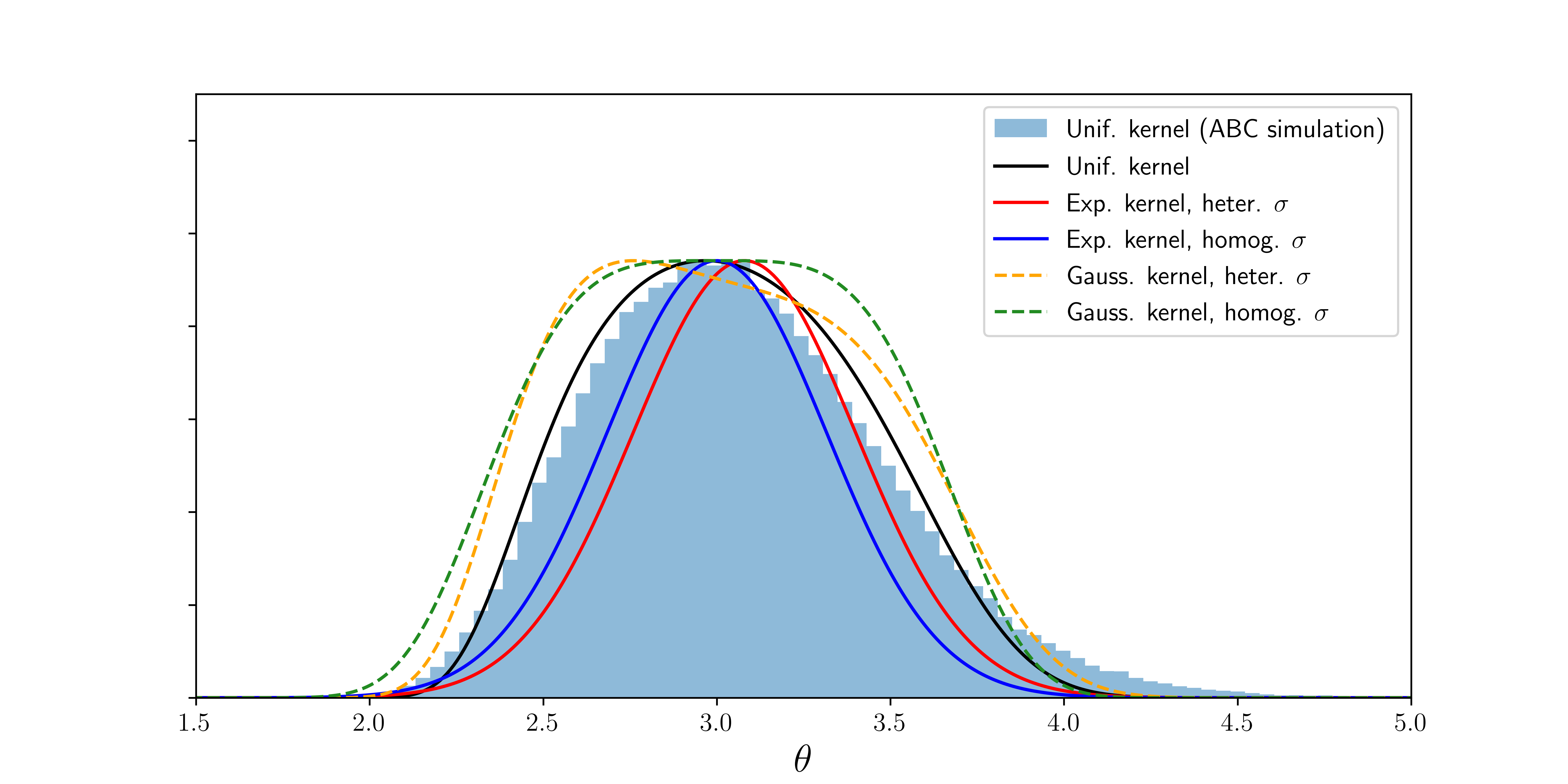}
\caption{Illustration of the different approximate posterior distributions of the model defined in Example \ref{example:1} approximated based on the various kernel assumptions made in Section \ref{subsec:gaus_discr_analysis}.  } \label{fig:example_1}
\end{figure}

\end{exmp}

\subsection{The effect of transforming the discrepancy} \label{sec:add_analysis}

As mentioned, the Gaussian assumption in  \eqref{eq:gaus}, and consequently the connections derived above, hold only approximately in practice. 
%
Sometimes it could be more appropriate to assume instead that $\log\Delta_{\Btheta}$ is Gaussian distributed at each $\Btheta$. This alternative corresponds to the assumption of log-Normal distribution of the discrepancy at each $\Btheta$ and the above connections hold with $\log\Delta_{\Btheta}$ in the place of $\Delta_{\Btheta}$ in this case. Obviously, many other transformations could also be used. 
It is however not immediately clear how such transformation affects the meaningfulness and interpretation of the resulting ABC posterior. 
We next discuss this in the case of a strictly increasing transformation $g:\realsp\rightarrow\reals$.

We first observe that $\prob(\Delta_{\Btheta}\leq h) = \prob(g(\Delta_{\Btheta})\leq h')$ where $h'=g(h)$. This implies that the use of the transformed discrepancy $g(\Delta_{\Btheta})$ does not alter the ABC target posterior based on the uniform kernel as long as the threshold $h$ is similarly transformed to $g(h)$. 
Nevertheless, the transformation affects the ABC posterior based on a non-uniform kernel or \kernel{}. Clearly, the generalized posterior $\genpostgd(\Btheta\cond\Bxobs)$ also depends on $g$ in analogy how different transformations of the loss function in Bayesian decision theory lead to different estimators (e.g.~the quadratic loss produces the mean whereas its square root, the absolute error, leads to the median, see \citet[Section~2.5]{Robert2007} for details). Namely, if $\Delta'_{\Btheta}=g(\Delta_{\Btheta})$ at each $\Btheta$, then
%
\begin{equation}
    \int K_h(\Delta_{\Btheta})\pdf(\Delta_{\Btheta}) \ud\Delta_{\Btheta} = \int K_h(g^{-1}(\Delta'_{\Btheta}))\pdf(\Delta'_{\Btheta}) \ud\Delta'_{\Btheta}, \label{eq:transf_K}
\end{equation}
which shows that the \kernel{} $K'_h(r) \eqdef K_h(g^{-1}(r))$ for the transformed discrepancy $\Delta'_{\Btheta}$ corresponds the original kernel or \kernel{} $K_h(r)$. 

For example, suppose $\Delta'_{\Btheta}=\log\Delta_{\Btheta}$ and that the Exponential \kernel{} $K'_h(r)\propto\e^{-r/h}$ is placed for $\Delta'_{\Btheta}$. We can now see that this is equivalent to using the \kernel{} 
%
\begin{equation}
    K_h(r) \propto \e^{-\log(r)/h} = r^{-1/h} \label{eq:exp_log_K}
\end{equation}
for the original, non-negative discrepancy $\Delta_{\Btheta}$. The \kernel{} $K_h(r) \propto r^{-1/h}$ is neither bounded at $r=0$ nor integrable with any $h>0$ and it hence cannot be interpreted as a probability density. The use of this \kernel{} for ABC may still be allowed whenever the resulting approximate posterior is a proper density (which though may be difficult to assess in practice), in analogy how an improper prior is commonly accepted in standard Bayesian inference as long as the resulting posterior is proper. 

\citet{Schmon2021} proposed generalized ABC likelihood 
\begin{equation}
    \pdfabcl(\Bxobs \cond \Btheta) = \int \e^{-wl(\Bxobs,\Bx)} \pdf(\Bx\cond\Btheta) \ud\Bx, \label{eq:schmon_genabc}
\end{equation}
which is based on a loss function $l:\X^2\rightarrow\reals$. A motivation for the loss $l(\Bxobs,\Bx)$ is that the error model $\pdf(\Bxobs\cond\Bx)$ for $\pdfabcerr(\Bxobs \cond \Btheta)$ in  \eqref{eq:abc_lik_errmodel} would often be difficult to specify correctly. Clearly, if $l(\Bxobs,\Bx)=-\log\pdf(\Bxobs\cond\Bx)$ and $w=1$ then $\pdfabcl(\Bxobs \cond \Btheta)$ simplifies to $\pdfabcerr(\Bxobs \cond \Btheta)$. 
Now, we may regard the ABC likelihood with the \kernel{} $K_h(r) \propto r^{-1/h}$ in  \eqref{eq:exp_log_K} as a special case of this generalized ABC when $l(\Bxobs,\Bx)=\log\Delta(\Bxobs,\Bx)$ and $w=1/h$ whereas the original Exponential kernel would be obtained using $l(\Bxobs,\Bx)=\Delta(\Bxobs,\Bx)$ and $w=1/h$. 
Note that the approach by \citet{Schmon2021} is in general not equivalent to ours because there the loss function is used in the place of the negative log-likelihood of the observation model $-\log\pdf(\Bxobs\cond\Bx)$ whereas we consider replacing the intractable negative log-likelihood $-\log\pdf(\Bxobs\cond\Btheta)$ with a loss function. 

\begin{figure}[hbtp] 
\centering
\includegraphics[width=0.8\textwidth]{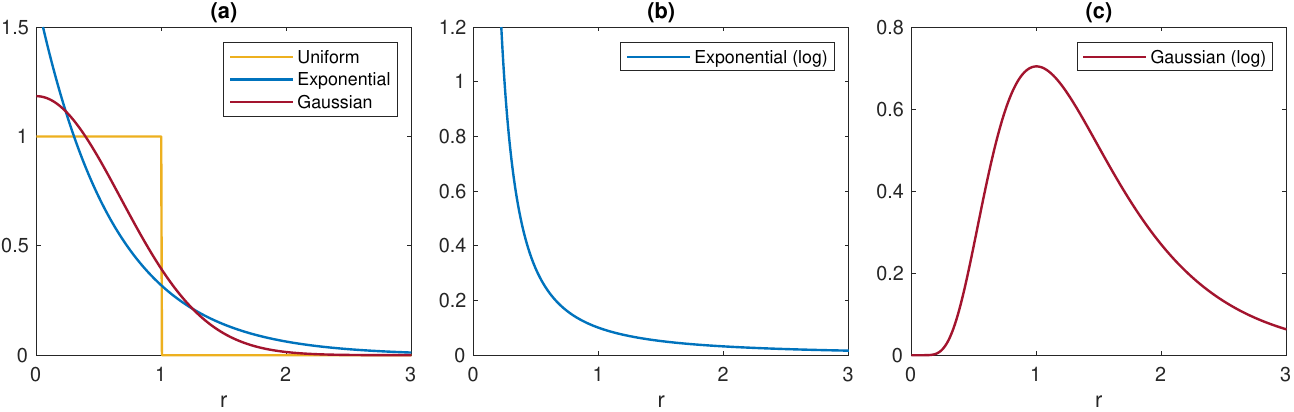}
\caption{(a) Uniform, Exponential and Gaussian kernels where $m_h=0$. (b) The \kernel{} in  \eqref{eq:exp_log_K} scaled to be roughly in correspondence with the Exponential kernel of (a). (c) The \kernel{} in  \eqref{eq:gaus_log_K} with $m_h=0$.} \label{fig:demo_kernels1}
\end{figure}

If $\Delta'_{\Btheta}=\log\Delta_{\Btheta}$ and the \kernel{} for $\Delta'_{\Btheta}$ is $K'_h(r) \propto \e^{-(r-m_h)^2/(2\sigma_h^2)}$, then the equivalent \kernel{} for $\Delta_{\Btheta}$ is $K_h(r) \propto \e^{-(\log(r)-m_h)^2/(2\sigma_h^2)}$. It is easy to check that this function is integrable and hence defines a proper density. It can be normalized so that  
%
\begin{equation}
    K_h(r) = \frac{1}{\e^{m_h+\sigma_h^2/2}\sqrt{2\pi\sigma_h^2}}\e^{-(\log(r)-m_h)^2/(2\sigma_h^2)} \label{eq:gaus_log_K}
\end{equation}
%
for $r>0$. 
Hence,  \eqref{eq:gaus_log_K} could be interpreted as a probabilistic error model. Unlike with the other \kernel{}s encountered, we have $\lim_{r\rightarrow 0^{+}}K_h(r)=0$ and $\arg\max_{r>0}K_h(r)=\e^{m_h}>0$ with any $\sigma_h>0$ and $m_h\in\reals$. These properties assert that some systematic error or model misspecification is implicitly assumed in this case. 
Figure \ref{fig:demo_kernels1} illustrates the kernels and \kernel{}s that appear in this paper. 

An important remark is that the above connections also require that $h$ is not too small as the approximate nature of the Gaussian assumption becomes apparent in the lower tail otherwise. This is in line with the fact that the (exact) ABC target posterior $\pdfabc(\Btheta \cond \Bxobs)$ converges to the standard posterior $\pdf(\Btheta\cond\Bxobs)$ as $h\rightarrow 0$ (under certain technical conditions regarding the kernel, summary statistics and the discrepancy) but the generalized posterior $\genpostd(\Btheta\cond\Bxobs)$ does not satisfy any similar result. This disagreement is however irrelevant in practice where $h$ typically cannot be set negligible to keep the computational efficiency of ABC at a satisfactory level. Also, it may be impossible to simulate data that produces small discrepancy values anyway when the model is misspecified, see \citet[Section~2.1]{Frazier2020misabc} for some discussion. 
On the other hand, the ABC posterior and the generalized posterior are both equal to the prior when $h\rightarrow\infty$ and $w\rightarrow 0$. 

In this paper we only investigate how the generalized posterior and the target posterior of ABC are related when the same distance is involved in both. 
It is still worth noting that the current principles for forming the discrepancy for ABC and selecting the loss function for GBI differ and they serve different purpose. GBI has been studied mainly in the context of i.i.d.~data where the loss has an additive structure with respect to the data points so that $l(\Btheta,\Bxobs) = \sum_i l'(\Btheta,\Bxobsi)$ where $l'$ is some pointwise loss function. If the generalized posterior is updated sequentially in this case, the order of the data does not affect the resulting density because $\genpost(\Btheta\cond\Bxobs) \propto \pdf(\Btheta)\e^{-w\sum_i l'(\Btheta,\Bxobsi)} = \pdf(\Btheta) \prod_i \e^{-w l'(\Btheta,\Bxobsi)} \propto \pdf(\Btheta)\prod_i\genpostpr(\Btheta\cond\Bxobsi)$. This coherence property was also used by \citet{Bissiri2016} to justify the exponential form of $\genpost(\Btheta\cond\Bxobs)$ in \eqref{eq:genpost}. On the other hand, the discrepancy in ABC is commonly tailored for each computational task at hand and is often based on summary statistics so that coherence does not hold in general. 
Further, ABC is mainly used in a non-sequential data setting as a computational approximation for the exact posterior that satisfies the coherence property, so the lack of coherence may be considered somewhat irrelevant in this regard. 

Our final remark concerns the basic computational methods for the ABC and GBI posteriors considered. 
The ABC likelihoods $\pdfabc(\Bxobs \cond \Btheta)$ and $\pdfabcerr(\Bxobs \cond \Btheta)$ can be estimated unbiasedly using only model simulations from $\pdf(\Bx\cond\Btheta)$. This facilitates pseudo-marginal MCMC and other sampling algorithms that exactly target the ABC posterior but which may be computationally costly. The expected loss $l_{\Delta}(\Btheta,\Bxobs)$ can similarly be estimated but this leads to the unbiased estimate for the \emph{log-}density $\log\genpostd(\Bxobs\cond\Btheta) = -wl_{\Delta}(\Btheta,\Bxobs)$ so that it does not seem feasible to apply pseudo-marginal MCMC in an exact fashion. On the other hand, one can use various surrogate models to estimate the scalar-valued $l_{\Delta}(\Btheta,\Bxobs)$ and then use this estimate in $\genpostd(\Btheta\cond\Bxobs)$ which allows straightforward and computationally efficient approximate inference. For example, \citet{Gao2023} uses neural networks in an amortized setting 
whereas \citet{Thomas2023,Jarvenpaa2023} use Gaussian process models, as is further discussed in the following section, in the case of a single set of observed data. 

\section{Examples} \label{sec:exp}\label{subsec:bact}

In this section we discuss the above connections in two related settings: 
First, in Section \ref{subsec:ex1}, we extend our analysis to explain the numerically observed similarity of the ABC and surrogate-based generalized posteriors in an inference problem with real data used in \citet[Section~7.3.2]{Jarvenpaa2023}. 
In Section \ref{subsec:ex2} we discuss the selection of the scaling constant $w$ in the context of the surrogate-based GBI algorithm for intractable models recently proposed by \citet{Thomas2023} which provides some new insight on their method. 

\subsection{Example 1: The observed similarity of particular GBI and ABC posteriors} \label{subsec:ex1}

We revisit the numerical results of \citet[Section~7.3.2]{Jarvenpaa2023} where a Gaussian process surrogate-based Metropolis-Hastings algorithm called GP-MH was briefly demonstrated with a target density of the form $\genpostd(\Btheta\cond\Bxobs)$. 
This target density was considered as an alternative to ABC for estimating the three parameters of a statistical model by \citet{Numminen2013} which describes the transmission dynamics of bacterial infections in day care centres. 
The unknown model parameters $\Btheta$ consist of the internal, external and co-infection parameters $\beta,\Lambda$ and $\theta$, respectively. The real data used for inference records colonizations with the bacterium \emph{Streptococcus pneumoniae} at $29$ day care centres. The discrepancy $\Delta$ was formed as the square root of the average of four $L^1$ distances defined between certain simulated and observed summary statistics and empirically computed over the $29$ day care centres. 
The GP surrogate model in the GP-MH algorithm facilitates sample-efficient approximation of $\Edth$ for the target density. The surrogate modeling is here important because the bacterial infections model is rather costly to simulate and its likelihood function is unavailable. 
Further details of the model and the inference procedure can be found in \citet{Numminen2013} and \citet{Jarvenpaa2023}, respectively. 

The numerical results of \citet{Jarvenpaa2023} showed that in this case the choice 
\begin{equation}
    w = 20 \approx \frac{2}{\Sd(\Delta_{\Btheta_*})}, 
    \label{eq:wchoice}
\end{equation}
where $\Btheta_*$ is the parameter that (approximately) minimizes $\Edth$ and where $\Sd(\Delta_{\Btheta_*}) = \sqrt{\Var(\Delta_{\Btheta_*})} = \sqrt{\Var_{\Bx\cond\Btheta_*}(\Delta(\Bxobs,\Bx))}$, resulted a generalized posterior similar to their ABC posterior that they computed for comparison. 
In the light of our analysis above, we can now explain this observation. First we must acknowledge that \citet{Jarvenpaa2023} used the uniform kernel with threshold $\epsi$ for ABC while our main connection is based on the Exponential kernel (\kernel{}). We can match the Exponential kernel 
to the uniform kernel by solving the optimization problem
\begin{equation}
    \min_{h>0} g_{\epsi}(h), \label{eq:unif_exp_matching}
\end{equation}
where
\begin{equation}
    g_{\epsi}(h) \eqdef \int_{0}^{\infty} \left| \frac{1}{\epsi}\indic_{\Delta\leq\epsi} - \frac{1}{h}\e^{-\Delta/h} \right| \ud\Delta.
\end{equation}
We first observe that if $\epsi\leq h$, then
\begin{align}
    g_{\epsi}(h) 
    = \int_{0}^{\epsi} \left(\frac{1}{\epsi}-\frac{1}{h}\e^{-\Delta/h} \right) \ud\Delta 
    + \int_{\epsi}^{\infty}\frac{1}{h}\e^{-\Delta/h} \ud\Delta 
    = 2\e^{-\epsi/h},
\end{align}
which is clearly minimized when $h=\epsi$. The solution is hence satisfies $h \leq \epsi$. 
On the otherhand, if $0 < h \leq \epsi$, then
\begin{align}
    g_{\epsi}(h) 
    &= \int_{0}^{b}\left(\frac{1}{h}\e^{-\Delta/h} -\frac{1}{\epsi}\right) \ud\Delta 
    + \int_{b}^{\epsi} \left(\frac{1}{\epsi}-\frac{1}{h}\e^{-\Delta/h} \right) \ud\Delta
    + \int_{\epsi}^{\infty} \frac{1}{h}\e^{-\Delta/h} \ud\Delta \\
    &= 2\left( \frac{\epsi-h-b}{\epsi} + \e^{-\epsi/h} \right),
\end{align}
where $b \eqdef h\log(\epsi/h)$ is the solution to $1/\epsi = (1/h)\e^{-b/h}$. 
We denote $0<a \eqdef h/\epsi \leq 1$ so that
\begin{align}
    g(a) &\eqdef g_{\epsi}(h) = 2\left( 1 - a + a\log(a) + \e^{-1/a} \right), \\
    g'(a) &= \frac{\partial g(a)}{\partial a} = 2\frac{a^2\log(a) + \e^{-1/a}}{a^2} \label{eq:dgda} \\
    g''(a) &= \frac{\partial^2 g(a)}{\partial a^2} = \frac{\e^{-1/a} \left( a^3 \e^{1/a} - 2a + 1 \right)}{a^4}
\end{align}
Because $\lim_{a\rightarrow 0^+}g'(a) = -\infty$ (which follows from the fact $\lim_{a\rightarrow 0^+}\e^{-1/a}/a^2 = \lim_{c\rightarrow \infty}c^2/\e^{c}=0$), $g'(1) = 2/\e>0$ and because $g'$ is continuous in $(0,1)$, it has a root in $(0,1)$. As the second derivative of $g(a)$ is positive in $(0,1)$ the root is unique, gives the global minimum and can be numerically approximated as $h/\epsi \approx 0.6 \Leftrightarrow h \approx 0.6\epsi$.

However, the ABC posterior with the Exponential \kernel{} is invariant to any translation of $\Delta$ because $\e^{-(\Delta_{}-c)/h} \propto \e^{-\Delta_{}/h}$ for any constant $c$ but the one with uniform \kernel{} is not unless the threshold $\epsi$ is also translated. 
Hence, we consider ``standardized'' discrepancy obtained by replacing $\Delta(\Bxobs,\Bx)$ with $\Delta(\Bxobs,\Bx)-\Delta_0$ where $\Delta_0\geq 0$ is the smallest possible discrepancy value that can be simulated. Due to model misspecification, $\Delta_0$ is not necessarily zero. In practice, $\Delta_0$ needs to be estimated numerically and could alternatively be defined as the smallest discrepancy value obtainable with non-negligible probability in some sense. 
That is, we replace $\epsi$ with $\epsilon \eqdef \epsi-\Delta_0$ in  \eqref{eq:unif_exp_matching} and treat the ABC posterior based on the Exponential kernel with $h \approx 0.6\epsilon = 0.6(\epsi-\Delta_0)$ as an approximation to the ABC posterior based on the uniform kernel with threshold $\epsi$. 

Now, if $\Delta_{\Btheta} \sim \Normal(\Edth,\Vdth)$ and we take $\Delta_0$ to be, say, the 95\% lower tail probability at $\Btheta = \Btheta_* \eqdef \arg\min_{\Btheta\in\Theta}\Edth$ (which is here assumed to be unique), then $\Delta_0 \approx \Edthstar - 1.96\,\Sdthstar$. If $\epsi = \min_{\Btheta\in\Theta}\Edth$, as used as the default setting in some implementations of GP-based ABC algorithms in the ELFI software \citep{Lintusaari2017}, 
then $\epsilon \approx 1.96\,\Sdthstar$. 
We then have $w = 1/h \approx 1/(0.6\epsilon) \approx 0.9/\Sdthstar$. If we instead use the more conservative choice $\epsi = \Edthstar-\Sdthstar$, we similarly see that $w = 1/h \approx 1.7/\Sdthstar$ which approximately matches  \eqref{eq:wchoice}. 

Specifically, the Gaussianity assumption of the discrepancy in the Bacterial infections example holds approximately though $\Vdth$ is not exactly constant but appears to grow as a function of the magnitude of the mean discrepancy. We numerically see that $\Delta_{\Btheta_*} \approx \Normal(1.1, 0.11^2)$ and the threshold in the ABC baseline of \citet{Jarvenpaa2023} was set to $\epsi=1.0$. As above we obtain $\Delta_0 = 1.1 - 1.96 \cdot 0.11 = 0.88$ so that $\epsilon \approx 0.12$ and $w \approx 1/(0.6\epsilon) \approx 14$ which is not far from  \eqref{eq:wchoice}. This analysis explains the observed similarity of the generalized and ABC posteriors in this particular example and shows that the generalized posterior can indeed be treated as an approximation to the ABC target itself in some problems. 

Nevertheless, under the Gaussianity assumption, the loss in  \eqref{eq:exploss1} based on the Exponential \kernel{} grows linearly with respect to $\Edth$ whereas the ``ABC loss'' in  \eqref{eq:log_gauss_abc_loss} based on the uniform \kernel{} grows faster than linearly for large $\Edth$. Consequently, even if the densities are here similar, especially considering the various inevitable computational approximations involved, the tails are in general likely heavier in the former case. This can be seen as a consequence of more robust Bayesian inference. 

\subsection{Example 2: On the selection of scaling constant $w$ in \citet{Thomas2023} } \label{subsec:ex2}

\citet{Thomas2023} propose an approach towards addressing the very challenging task of likelihood-free inference in high-dimensional parameter space. Their algorithm is also based on Gaussian process surrogate modeling. They also acknowledge model misspecification in this setting and, although this aspect is somewhat nonrigorously presented in their paper, their ultimate target density can be considered to be of the form $\genpostd(\Btheta\cond\Bxobs)$. 
They assert that the discrepancy can be meaningfully formed to have an additive structure with respect to a partition of the parameter $\Btheta$ into some $q$ disjoint components and so that each additive term of the resulting expected loss has separate scaling factor $1/\delta_i, i=1,\ldots,q$. 
Here we disregard the additivity condition, as it is mainly used to tackle the challenge of high-dimensional parameter, and briefly discuss the selection of $\delta=\delta_1$ in the important special case of $q=1$ in the light of our analysis. 

When translated to our set-up, the suggestion by \citet{Thomas2023} is to select $\delta$ in practice as
\begin{equation}
    \delta = \max\Big\{ \min_{\Btheta\in\Theta} {\mean}(\Delta_{\Btheta}), \min_{i}\{\Delta^{(i)}\} \Big\},
    \label{eq:delta}
\end{equation}
where $\Delta^{(i)}$ denote the discrepancies calculated from the simulator queries while running the algorithm. In practice the exact value $\mean(\Delta_{\Btheta})$ in  \eqref{eq:delta} of course needs to be replaced by some estimate obtained e.g.~by using some GP surrogate model or neural network. The maximum in  \eqref{eq:delta} ensures that $\delta$ is non-negative, as all $\Delta^{(i)}$ will be non-negative.

\citet{Thomas2023} use $1/\delta$ in place of our $w$ in $\genpostd(\Btheta\cond\Bxobs)$ so that when $w=1/h$, we have $h=\delta$. Hence, if the discrepancy is Gaussian distributed with constant variance, which is the key assumption of their GP-surrogate approach, we observe that their ultimate GBI target density is approximately equal to the ABC posterior based on the Exponential kernel with bandwidth $h=\delta$ given by  \eqref{eq:delta}. 

To be able to better interpret the selection $h=\delta$, we may apply the idea of matching the uniform kernel to the Exponential one as in Section \ref{subsec:ex1}. We then obtain $\delta \approx 0.6(\epsi-\Delta_0)$ so that $\epsi \approx 1.7\delta + \Delta_0$. In the typical case where  \eqref{eq:delta} simplifies to $\delta = \min_{\Btheta\in\Theta} {\mean}(\Delta_{\Btheta})$, this choice of $\epsi$ would be conservative for ABC irrespective of how one deals with the selection of $\Delta_0\geq 0$ (e.g.~in Section \ref{subsec:ex1} we had $\epsi \leq \min_{\Btheta\in\Theta} {\mean}(\Delta_{\Btheta})$). Namely, if the variance of the discrepancy is small as compared to the mean, then almost all model simulations could be accepted near $\arg\min_{\Btheta\in\Theta} {\mean}(\Delta_{\Btheta})$ which usually means inflated uncertainty quantification as compared to the exact posterior. On the other hand, if the model is crudely misspecified then this behavior could be beneficial. 
The goal of this paper is however not to investigate model misspecification in particular.

\section{Conclusions} \label{sec:concl}

Generalized Bayesian inference has attracted more interest during the recent years with an increasing level of awareness of the undesirable behavior of posterior distributions under model misspecification and large data sets. Here we focused on scrutinizing the relationship between generalized posterior $\genpostd(\Btheta\cond\Bxobs)$ with the expected discrepancy $l_{\Delta}(\Btheta,\Bxobs)$ as the loss function and the ABC target posterior $\pdfabc(\Bxobs \cond \Btheta)$ with discrepancy $\Delta(\Bxobs,\Bx)$ in situations of practical relevance. 
This connection thus offers an alternative interpretation of the generalized posterior, which has been used as a target density of some recently proposed approximate inference algorithms such as \citet{Jarvenpaa2023,Gao2023,Thomas2023}, as a justifiable approximation to the target posterior of ABC. 

A potential challenge with the mentioned connection, however, is that it might be difficult to assess in practice whether the GBI posterior defines an alternative approach to ABC due to the Gaussian assumption involved. 
We did not examine here whether the connection can be used to achieve significant computational improvements over existing ABC methods, other than emphasizing how it allows easier use of surrogate-based approximate inference methods for intractable and computationally intensive models. 
Both of these aspects therefore warrant further investigation. 
Another potential area for further research would be to try to identify other useful links between generalized Bayesian inference and likelihood-free inference methods that may help to design computationally more efficient or more robust inference algorithms. 



\subsubsection*{Acknowledgments}

This research was funded by Research Council of Norway, grant no. 299941 and through its Centre of Excellence Integreat - The Norwegian Centre for knowledge-driven machine learning, project number 332645 and by the European Research Council grant no.~742158.


\appendix
\numberwithin{equation}{section}
\numberwithin{figure}{section}

\bibliography{references}

\begin{thebibliography}{26}
\providecommand{\natexlab}[1]{#1}
\providecommand{\url}[1]{\texttt{#1}}
\expandafter\ifx\csname urlstyle\endcsname\relax
  \providecommand{\doi}[1]{doi: #1}\else
  \providecommand{\doi}{doi: \begingroup \urlstyle{rm}\Url}\fi

\bibitem[Bissiri et~al.(2016)Bissiri, Holmes, and Walker]{Bissiri2016}
P.~G. Bissiri, C.~C. Holmes, and S.~G. Walker.
\newblock A general framework for updating belief distributions.
\newblock \emph{Journal of the Royal Statistical Society: Series B (Statistical
  Methodology)}, 78\penalty0 (5):\penalty0 1103--1130, 2016.

\bibitem[David T.~Frazier and Koo(2024)]{Frazier2022lossbased}
Gael M.~Martin David T.~Frazier, Rubén Loaiza-Maya and Bonsoo Koo.
\newblock Loss-based variational bayes prediction.
\newblock \emph{Journal of Computational and Graphical Statistics}, 0\penalty0
  (0):\penalty0 1--12, 2024.
\newblock \doi{10.1080/10618600.2024.2341899}.
\newblock URL \url{https://doi.org/10.1080/10618600.2024.2341899}.

\bibitem[Frazier et~al.(2020)Frazier, Robert, and Rousseau]{Frazier2020misabc}
D.~T. Frazier, C.~P. Robert, and J.~Rousseau.
\newblock {Model Misspecification in Approximate Bayesian Computation:
  Consequences and Diagnostics}.
\newblock \emph{Journal of the Royal Statistical Society Series B: Statistical
  Methodology}, 82\penalty0 (2):\penalty0 421--444, 01 2020.

\bibitem[Gao et~al.(2023)Gao, Deistler, and Macke]{Gao2023}
Richard Gao, Michael Deistler, and Jakob~H. Macke.
\newblock Generalized bayesian inference for scientific simulators via
  amortized cost estimation.
\newblock In \emph{Thirty-seventh Conference on Neural Information Processing
  Systems}, 2023.
\newblock URL \url{https://openreview.net/forum?id=ZARAiV25CW}.

\bibitem[Gr{\"u}nwald and van Ommen(2017)]{Grunwald2017}
P.~Gr{\"u}nwald and T.~van Ommen.
\newblock {Inconsistency of Bayesian Inference for Misspecified Linear Models,
  and a Proposal for Repairing It}.
\newblock \emph{Bayesian Analysis}, 12\penalty0 (4):\penalty0 1069--1103, 2017.

\bibitem[Gutmann and Corander(2016)]{Gutmann2016}
M.~U. Gutmann and J.~Corander.
\newblock {Bayesian optimization for likelihood-free inference of
  simulator-based statistical models}.
\newblock \emph{Journal of Machine Learning Research}, 17\penalty0
  (125):\penalty0 1--47, 2016.

\bibitem[Jewson et~al.(2018)Jewson, Smith, and Holmes]{Jewson2018}
J.~Jewson, J.~Q. Smith, and C.~Holmes.
\newblock {Principles of Bayesian Inference Using General Divergence Criteria}.
\newblock \emph{Entropy}, 20\penalty0 (6), 2018.

\bibitem[Järvenpää and Corander(2024)]{Jarvenpaa2023}
M.~Järvenpää and J.~Corander.
\newblock {Approximate Bayesian inference from noisy likelihoods with Gaussian
  process emulated MCMC}.
\newblock \emph{Journal of Machine Learning Research}, 25\penalty0
  (366):\penalty0 1--55, 2024.

\bibitem[Järvenpää et~al.(2019)Järvenpää, Gutmann, Pleska, Vehtari, and
  Marttinen]{Jarvenpaa2018_acq}
M.~Järvenpää, M.~U. Gutmann, A.~Pleska, A.~Vehtari, and P.~Marttinen.
\newblock Efficient acquisition rules for model-based approximate {B}ayesian
  computation.
\newblock \emph{Bayesian Analysis}, 14\penalty0 (2):\penalty0 595--622, 2019.

\bibitem[Järvenpää et~al.(2020)Järvenpää, Vehtari, and
  Marttinen]{Jarvenpaa2020_babc}
M.~Järvenpää, A.~Vehtari, and P.~Marttinen.
\newblock {Batch simulations and uncertainty quantification in Gaussian process
  surrogate approximate Bayesian computation}.
\newblock In \emph{Proceedings of the 36th Conference on Uncertainty in
  Artificial Intelligence (UAI)}, pages 779--788, 2020.

\bibitem[Knoblauch et~al.(2022)Knoblauch, Jewson, and Damoulas]{Knoblauch2022}
J.~Knoblauch, J.~Jewson, and T.~Damoulas.
\newblock {An Optimization-centric View on Bayes' Rule: Reviewing and
  Generalizing Variational Inference}.
\newblock \emph{Journal of Machine Learning Research}, 23\penalty0
  (132):\penalty0 1--109, 2022.

\bibitem[Lintusaari et~al.(2018)Lintusaari, Vuollekoski,
  Kangasr\"{a}\"{a}si\"{o}, Skyt\'{e}n, J\"{a}rvenp\"{a}\"{a}, Gutmann,
  Vehtari, Corander, and Kaski]{Lintusaari2017}
J.~Lintusaari, H.~Vuollekoski, A.~Kangasr\"{a}\"{a}si\"{o}, K.~Skyt\'{e}n,
  M.~J\"{a}rvenp\"{a}\"{a}, M.~U. Gutmann, A.~Vehtari, J.~Corander, and
  S.~Kaski.
\newblock {ELFI: Engine for Likelihood Free Inference}.
\newblock \emph{Journal of Machine Learning Research}, 19\penalty0
  (16):\penalty0 1--7, 2018.

\bibitem[Matsubara et~al.(2022)Matsubara, Knoblauch, Briol, and
  Oates]{Matsubara2022}
T.~Matsubara, J.~Knoblauch, F.-X. Briol, and C.~J. Oates.
\newblock {Robust Generalised Bayesian Inference for Intractable Likelihoods},
  2022.
\newblock Available at \url{https://arxiv.org/abs/2104.07359}.

\bibitem[Numminen et~al.(2013)Numminen, Cheng, Gyllenberg, and
  Corander]{Numminen2013}
E.~Numminen, L.~Cheng, M.~Gyllenberg, and J.~Corander.
\newblock Estimating the transmission dynamics of streptococcus pneumoniae from
  strain prevalence data.
\newblock \emph{Biometrics}, 69\penalty0 (3):\penalty0 748--757, 2013.

\bibitem[Oliveira et~al.(2021)Oliveira, Ott, and Ramos]{Oliveira2021}
R.~Oliveira, L.~Ott, and F.~Ramos.
\newblock {No-regret approximate inference via Bayesian optimisation}.
\newblock In \emph{Proceedings of the Thirty-Seventh Conference on Uncertainty
  in Artificial Intelligence}, volume 161, pages 2082--2092, 2021.

\bibitem[Pacchiardi et~al.(2024)Pacchiardi, Khoo, and Dutta]{Pacchiardi2021}
Lorenzo Pacchiardi, Sherman Khoo, and Ritabrata Dutta.
\newblock {Generalized Bayesian likelihood-free inference}.
\newblock \emph{Electronic Journal of Statistics}, 18\penalty0 (2):\penalty0
  3628 -- 3686, 2024.
\newblock \doi{10.1214/24-EJS2283}.
\newblock URL \url{https://doi.org/10.1214/24-EJS2283}.

\bibitem[Pesonen et~al.(2023)Pesonen, Simola, Köhn-Luque, Vuollekoski, Lai,
  Frigessi, Kaski, Frazier, Maneesoonthorn, Martin, and Corander]{Pesonen2023}
H.~Pesonen, U.~Simola, A.~Köhn-Luque, H.~Vuollekoski, X.~Lai, A.~Frigessi,
  S.~Kaski, D.~T. Frazier, W.~Maneesoonthorn, G.~M. Martin, and J.~Corander.
\newblock {ABC of the future}.
\newblock \emph{International Statistical Review}, 91\penalty0 (2):\penalty0
  243--268, 2023.

\bibitem[Robert(2007)]{Robert2007}
C.~P Robert.
\newblock \emph{The Bayesian Choice}.
\newblock Springer, New York, second edition, 2007.

\bibitem[Schmon et~al.(2021)Schmon, Cannon, and Knoblauch]{Schmon2021}
S.~M. Schmon, P.~W. Cannon, and J.~Knoblauch.
\newblock {Generalized Posteriors in Approximate Bayesian Computation}.
\newblock In \emph{Third Symposium on Advances in Approximate Bayesian
  Inference}, 2021.

\bibitem[Sisson et~al.(2019)Sisson, Fan, and Beaumont]{Sisson2019}
S.~Sisson, Y.~Fan, and M.~Beaumont.
\newblock \emph{{Handbook of Approximate Bayesian Computation}}.
\newblock {New York: Chapman and Hall/CRC}, 2019.

\bibitem[Thomas et~al.(2023)Thomas, Sá-Leão, de~Lencastre, Kaski, Corander,
  and Pesonen]{Thomas2023}
O.~Thomas, R.~Sá-Leão, H.~de~Lencastre, S.~Kaski, J.~Corander, and
  H.~Pesonen.
\newblock Misspecification-robust likelihood-free inference in high dimensions,
  2023.
\newblock Available at \url{https://arxiv.org/abs/2002.09377}.

\bibitem[Walker(2013)]{Walker2013}
S.~G. Walker.
\newblock Bayesian inference with misspecified models.
\newblock \emph{Journal of Statistical Planning and Inference}, 143\penalty0
  (10):\penalty0 1621--1633, 2013.

\bibitem[Wilkinson(2013)]{Wilkinson2013}
R.~D. Wilkinson.
\newblock {Approximate Bayesian computation (ABC) gives exact results under the
  assumption of model error}.
\newblock \emph{Statistical Applications in Genetics and Molecular Biology},
  12\penalty0 (2):\penalty0 129--141, 2013.

\bibitem[Wu and Martin(2023)]{Wu2023}
P.-S. Wu and R.~Martin.
\newblock {A Comparison of Learning Rate Selection Methods in Generalized
  Bayesian Inference}.
\newblock \emph{Bayesian Analysis}, 18\penalty0 (1):\penalty0 105--132, 2023.

\bibitem[Yang(2007)]{Yang2007}
Z.~Yang.
\newblock {Fair-balance paradox, star-tree paradox, and Bayesian
  phylogenetics}.
\newblock \emph{Molecular Biology and Evolution}, 24:\penalty0 1639–1655,
  2007.

\bibitem[Yang and Zhu(2018)]{Yang2018}
Z.~Yang and T.~Zhu.
\newblock {Bayesian selection of misspecified models is overconfident and may
  cause spurious posterior probabilities for phylogenetic trees}.
\newblock \emph{Proceedings of the National Academy of Sciences}, 115:\penalty0
  1854--1859, 2018.

\end{thebibliography}
\bibliographystyle{plainnat}

\end{document}